\documentclass{appolb}
\usepackage[latin9]{inputenc}
\usepackage{amsmath}
\usepackage{graphicx}
\usepackage{esint}

\makeatletter

\providecommand{\tabularnewline}{\\}

\usepackage{graphicx}
\makeatother

\begin{document}

\title{An unitarized model for tetraquarks with a color flip-flip potential }

\author{Marco Cardoso, Pedro Bicudo \address{CFTP, Instituto Superior Técnico}
\\
}
\maketitle
\begin{abstract}
In this work, a color structure dependent flip-flop potential is developed
for the two quarks and two antiquarks system. Then, this potential
is applied to a microscopic quark model which, by integrating the
internal degrees of freedom, is transformed into a model of mesons
with non-local interactions. With this, the T matrix for the system
is constructed and meson-meson scattering is studied. Tetraquarks
states, interpreted as poles of of the T matrix, both bound states
and resonances, are found. Special emphasis is given to the truly
exoti\={c} $qq\bar{Q}\bar{Q}$ system, but some results for the crypto-exotic
$qQ\bar{q}\bar{Q}$ are also presented. 
\end{abstract}

\section{Introduction}

The existence of composite particles constituted by two quarks and
two antiquarks, tetraquarks, is still debated. Although several experimental
candidates \cite{Aaij:2014jqa,Ali:2009es} have been advanced no one
has been firmly established. From the theoretical point of view, these
systems were studied mainly as a bound state of two quarks and two
antiquarks \cite{Vijande:2007ix,Bicudo:2012qt}.

In this work we start with a microscopic model of two quarks and two
antiquarks interacting through a four-body potential. By integrating
the confined degrees of freedom we obtain a multi-channel model of
mesons. This model is then used to find bound states and to construct
the scattering T matrix, from were resonances are found.

\section{Method}

\subsection{Microscopic potential}

\begin{figure}
\begin{centering}
\includegraphics[width=0.2\columnwidth]{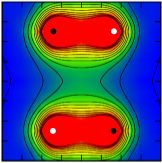}\includegraphics[width=0.2\columnwidth]{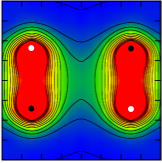}\includegraphics[width=0.2\columnwidth]{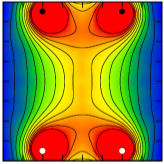}
\par\end{centering}

\protect\caption{The three possible string configurations for the ground state of a
system of two static quarks and two static antiquarks}
\label{Fig:TripleFF}
\end{figure}

The static potential has been found on the lattice \cite{Alexandrou:2004ak,Okiharu:2004ve}.
It is given by a triple flip-flop potential, where it's values corresponds
to the confining string disposition that minimizes the potential for
a given configuration (see Fig. \ref{Fig:TripleFF}): 
\begin{equation}
V_{FF}=\min(V_{I},V_{II},V_{T})
\end{equation}

$V_{I}$ and $V_{II}$ are the two-meson potentials
\begin{eqnarray}
V_{I} & = & V_{M}(r_{13})+V_{M}(r_{24})\\
V_{II} & = & V_{M}(r_{14})+V_{M}(r_{23})
\end{eqnarray}
where $V_{M}$ is the quark-antiquark potential in a meson, which
is well described by the Cornell potential $V_{M}=K-\frac{\gamma}{r}+\sigma\, r$.

$V_{T}$ is the tetraquark potential, given by
\[
V_{T}=2K-\gamma\sum_{i<j}\frac{C_{ij}}{r_{ij}}+\sigma\, L_{min}(\mathbf{x}_{1},\mathbf{x}_{2},\mathbf{x}_{3},\mathbf{x}_{4})
\]

where $C_{ij}=1/2$ between two quarks or two antiquarks and $C_{ij}=1/4$
between a quark and an antiquark. $L_{min}$ is the minimal length
of the string linking the four particles.

Two linearly independent color singlets can be formed from two quarks
and two antiquarks, say the two meson-meson states: $|\mathcal{C}_{I}\rangle=\frac{1}{3}|Q_{i}Q_{j}\overline{Q}_{i}\overline{Q}_{j}\rangle$
and$|\mathcal{C}_{II}\rangle=\frac{1}{3}|Q_{i}Q_{j}\overline{Q}_{j}\overline{Q}_{i}\rangle$,
or the color anti-symmetric and symmetric states $|\mathcal{A}\rangle=\frac{\sqrt{3}}{2}\big(|\mathcal{C}_{I}\rangle-|\mathcal{C}_{II}\rangle\big)$
and $|\mathcal{S}\rangle=\sqrt{\frac{3}{8}}\big(|\mathcal{C}_{I}\rangle+|\mathcal{C}_{II}\rangle\big)$.
We need a $2\times2$ matrix potential to be possible a transition
between the two states. So, we have to know the first excited potential
of the system, as well as the color structure of both states.

The color vector of the ground state could either be $|\mathcal{C}_{I}\rangle$
when $V_{FF}=V_{I}$, $|\mathcal{C}_{II}\rangle$ when $V_{FF}=V_{II}$or
$|\mathcal{A}\rangle$ when $V_{FF}=V_{T}$. As for the excited state,
we know it has to be orthogonal to the ground one since the potential
is hermitian. So we have $|\bar{\mathcal{C}}_{I}\rangle$ when $V_{FF}=V_{I}$,
$|\bar{\mathcal{C}}_{II}\rangle$ when $V_{FF}=V_{II}$and $|\mathcal{S}\rangle$
when $V_{FF}=V_{T}$, with $\langle\mathcal{C}_{A}|\bar{\mathcal{C}}_{A}\rangle=0$.
We assume that the value of the excited state is the second lowest
of the three potentials. This way we obtain the potential of the system.

\subsection{From Quarks to Mesons}

Since we study meson-meson interaction, the natural choice for the
color structure basis is the $|\mathcal{C}_{I}\rangle$ and $|\mathcal{C}_{II}\rangle$.
Note that in this basis $g_{AB}\equiv\langle\mathcal{C}_{A}|\mathcal{C}_{B}\rangle\neq\delta_{AB}$
\[
g=\begin{pmatrix}1 & \frac{1}{3}\\
\frac{1}{3} & 1
\end{pmatrix}
\]
Expanding the color states $\Psi=\Psi^{A}\mathcal{C}_{A}$, we arrive
at the Schrödinger equation
\begin{equation}
g_{AB}\hat{T}_{q}\Psi^{B}+\hat{V}_{AB}\Psi^{B}=Eg_{AB}\Psi^{B}
\end{equation}
Since we want a theory of mesons, we must have the kinetic energy
of both meson sectors, and not the kinetic energy of quarks $T_{I}=T_{q}+V_{I}\neq T_{II}=T_{q}+V_{II}\neq T_{q}$.
For this, we define the kinetic energy of meson in a way that is both
hermitian and gives the correct asymptotic states:
\[
\hat{T}_{S}=\left(\begin{array}{cc}
\hat{T}_{I} & \frac{\hat{T}_{I}+\hat{T}_{II}}{6}\\
\frac{\hat{T}_{I}+\hat{T}_{II}}{6} & \hat{T}_{II}
\end{array}\right)
\]

and
\[
\hat{V}_{S}=\left(\begin{array}{cc}
V_{11}-V_{I} & V_{12}-\frac{V_{I}+V_{II}}{6}\\
V_{12}-\frac{V_{I}+V_{II}}{6} & V_{22}-V_{II}
\end{array}\right)
\]

This gives a new Schrödinger equation with the same form. The components
$\Psi^{A}$ are then expanded in two meson states and so we obtain
the equation
\begin{equation}
\hat{T}_{\alpha\beta}\psi^{\beta}+\hat{V}_{\alpha_{\beta}}\psi^{\beta}=Eg_{\alpha\beta}\psi^{\beta}\label{eq:schro_mesons}
\end{equation}

where the greek letter index includes the color index $A$ and the
remaining quantum numbers index $i$. The potential $V$ has the form
\begin{align*}
\hat{V}_{AiAj}\psi^{Aj}= & V_{ij}(\mathbf{r})\psi^{Aj}(\mathbf{r})\\
\hat{V}_{AiBj}\psi^{Bj}= & \int d^{3}\mathbf{r}_{B}'\, v_{ij}(\mathbf{r}_{A},\mathbf{r}_{B}')\,\psi^{Bj}(\mathbf{r}_{B}')\qquad\mbox{when }A\ne B
\end{align*}
. $T_{\alpha\beta}$ and $g_{\alpha\beta}$ have similar structures.

\subsection{Asymptotic behavior}

Writing, each component as $\psi^{\alpha}(r)=\frac{u^{\alpha}(r)}{r}Y_{l_{\alpha}m_{\alpha}}$
, the asymptotic behavior of $u^{\alpha}(r)$ is 
\begin{equation}
u^{\alpha}(r)\rightarrow A_{i\alpha}\sqrt{\frac{\mu_{\alpha}}{k_{\alpha}}}\sin(k_{\alpha}r-\frac{l_{\alpha}\pi}{2}+\varphi_{i\alpha})+f_{i\alpha}e^{i(k_{\alpha}r-\frac{l_{\alpha}\pi}{2})}\label{eq:asymp}
\end{equation}
This leads to the definition of the scattering $T$ matrix for this
system
\begin{equation}
T_{ij}=\sum_{\alpha}\sqrt{\frac{k_{\alpha}}{\mu_{\alpha}}}A_{i\alpha}^{*}e^{-i\varphi_{i\alpha}}f_{j\alpha}\label{eq:tmatrix}
\end{equation}
To calculate it, we first generate $N_{open}$ eigenfunctions of the
$\hat{T}_{s}$ operator $\hat{T}_{S}\Psi_{0}=Eg\Psi_{0}$, where $N_{open}$is
the number of open channels. Then the base is orthogonalized with
the Gram-Schmidt procedure, using as inner product
\[
\langle\Psi_{0i}|\Psi_{0j}\rangle=\sum_{\alpha}A_{i\alpha}^{*}A_{j\alpha}\cos(\varphi_{i\alpha}-\varphi_{j\alpha})
\]
This product is a direct consequence of the asymptotic behavior Eq.
\ref{eq:asymp}. $A_{i\alpha}$ are computed by fitting the long range
behavior of the generated functions.

We calculate the $\Psi_{i}$ by solving Eq. \ref{eq:schro_mesons}
with $\Psi_{i}=\Psi_{0i}+\chi_{i}$ :
\[
(\hat{T}+\hat{V})\chi_{i}=Eg\chi_{i}-V\Psi_{0i}
\]
 From the long distance behavior of $\chi_{i}$ we find $f_{i\alpha}$
and calculate the $T$ matrix with Eq. \ref{eq:tmatrix}.

By continuing the definition of the $T$ matrix into the complex energy
plane we find it's poles which are tetraquark resonances.

\subsection{Bound states}

We need a very large box to be able to accurately find bound states,
if we use Dirichlet boundary conditions and the bound states have
a very small binding energy, having therefore a large spatial extension.
To solve Eq. \ref{eq:schro_mesons} using finite differences we employ
boundary conditions that depend on the energy
\[
\big[H+B(E)\big]u=Egu
\]

and try to find a zero on the determinant of the matrix $H+B(E)-Eg$.
Employing the Newton's method, it is found with the iteration
\[
E^{(n+1)}=E^{(n)}-\frac{1}{\mbox{Tr}[(H+B(E)-Eg)^{-1}(B'(E)-g)]}
\]
.

\section{Results}

In this work we neglect all spin and dynamical quark effects. The
meson kinematics is non-relativistic.

\subsection{Exotic channels}

For the exotic $qq\bar{Q}\bar{Q}$ system, we consider the wave-function
to be of the type
\[
\Psi=\Phi(\boldsymbol{\rho}_{13},\boldsymbol{\rho}_{24})\psi(\mathbf{r}_{13,24})\mathcal{C}_{I}+\xi\Phi(\boldsymbol{\rho}_{14},\boldsymbol{\rho}_{23})\psi(\mathbf{r}_{14,23})\mathcal{C}_{II}
\]
, where $\xi=\pm1$. This wavefunction includes space and color degrees
of freedom, but not spin. The functions $\Phi$ must have a definite
symmetry for the exchange of it's arguments: $\Phi(\mathbf{y},\mathbf{x})=s\Phi(\mathbf{x},\mathbf{y})$
with $s=\pm1$. This way, when we apply the exchange operators of
color and space $P_{ij}^{RC}$ we obtain
\begin{eqnarray*}
P_{12}^{RC}\Psi & = & \xi(-1)^{L_{r}}s\,\Psi\\
P_{34}^{RC}\Psi & = & \xi\,\Psi
\end{eqnarray*}

Including spin and since wave-function must be anti-symmetric for
quark and antiquark exchanges, we have $P_{12}\Psi=(-1)^{1+S_{12}}\xi(-1)^{L_{r}}s\Psi=-\Psi$
and $P_{34}\Psi=(-1)^{1+S_{34}}\xi\Psi=-\Psi$. In this work, we choose
$\xi=1$, $s(-1)^{L_{r}}=1$ and $L=0$. This gives, $S_{12}=S_{34}=0$
and so $S=0$. Consequently, we have $J=0$. We also choose states
of positive parity, only.

\begin{figure}
\centering{}\includegraphics[width=0.38\columnwidth]{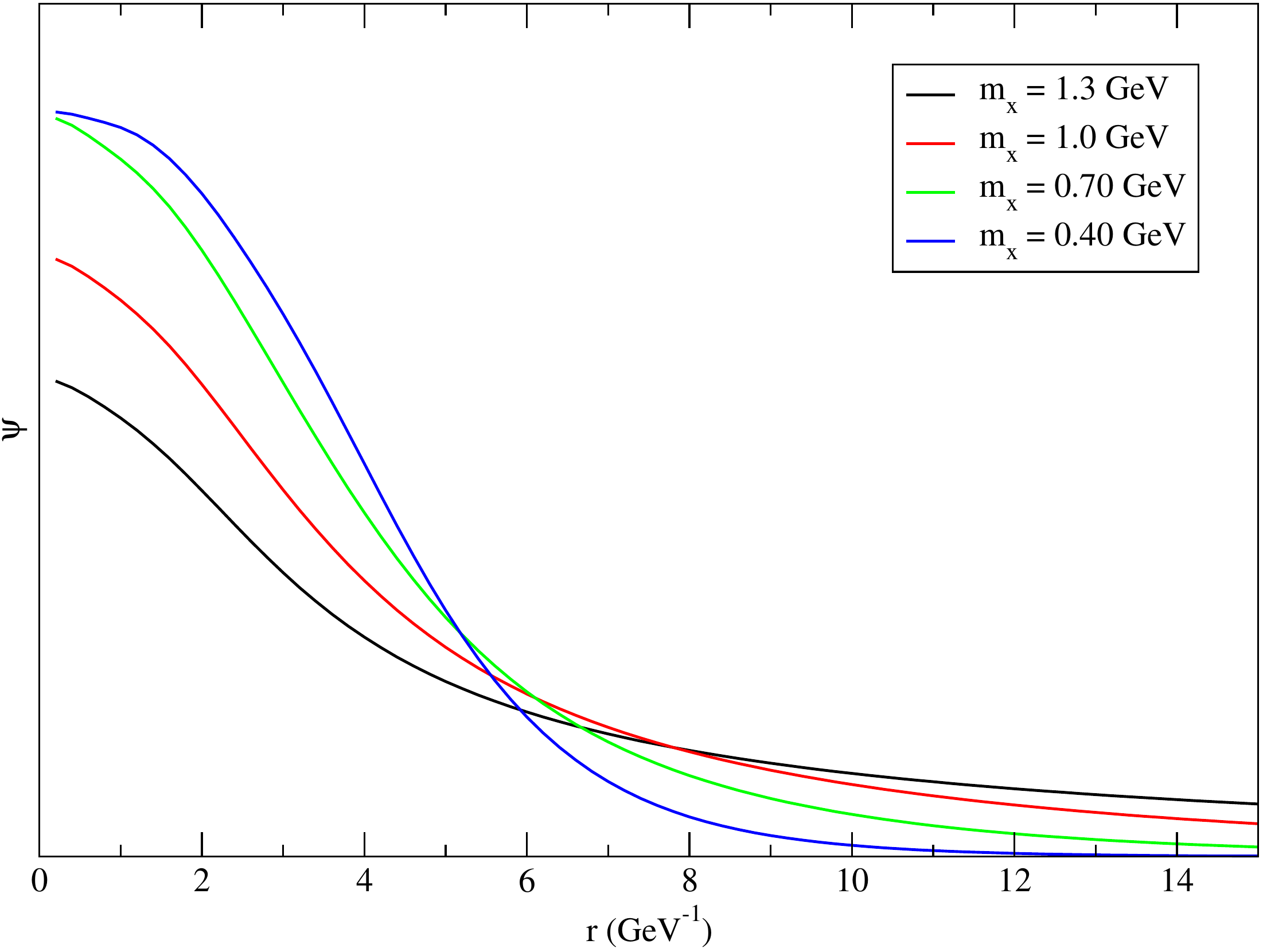}\protect\caption{Left: Bound state wavefunction for different masses of the lightest
quark in the $xx\bar{b}\bar{b}$ system. }
\label{Fig:exotic_plots}
\end{figure}

With $m_{\bar{Q}}=m_{b}=4.7\,\mbox{GeV}$, and varying the mass of
the quark from $m_{x}=0.40\,\mbox{GeV}$ to $m_{x}=1.3\,\mbox{GeV}$,
we find bound states for all the quark masses. Results for the binding
energy are given on table \ref{Table:boundstates} and the wave-functions
of the ground state component are shown in Fig. \ref{Fig:exotic_plots}.
For this system we find resonances between the opening of the the
second and third thresholds. Their complex energies are shown in Table
\ref{Table:resonances}.

Setting $m_{\bar{Q}}=1.3\,\mbox{GeV}$and similar quark masses, no
bound states or resonances are found.

\begin{table}
\begin{centering}
\begin{tabular}{|c|c|}
\hline 
$m_{x}\mathrm{(GeV)}$ & $B\mathrm{(MeV)}$\tabularnewline
\hline 
\hline 
1.30 & $\simeq0$\tabularnewline
\hline 
1.00 & -0.95\tabularnewline
\hline 
0.70 & -7.91\tabularnewline
\hline 
0.40 & -48.54\tabularnewline
\hline 
\end{tabular}
\par\end{centering}

\protect\caption{Binding energies of the $qq\bar{b}\bar{b}$ bound states for different
quark masses}
\label{Table:boundstates}
\end{table}

\begin{table}
\centering{}%
\begin{tabular}{|c|c|c|c|c|c|c|}
\cline{1-3} \cline{5-7} 
$m_{x}\mathrm{(GeV)}$ & $E\mathrm{(GeV)}$ & $N_{open}$ & \hspace{5mm} & $m_{x}\mathrm{(GeV)}$ & $E\mathrm{(GeV)}$ & $N_{open}$\tabularnewline
\cline{1-3} \cline{5-7} 
1.30 & 12.998 - 0.0179i & 2 &  & 0.70 & 11.545 - 0.237i & 1\tabularnewline
\cline{1-3} \cline{5-7} 
1.00 & 12.505 - 0.0192i & 2 &  &  & 12.019 - 0.033i & 2\tabularnewline
\cline{1-3} \cline{5-7} 
0.70 & 12.050 - 0.0215i & 2 &  & 0.40 & 11.431 - 0.024i & 1\tabularnewline
\cline{1-3} \cline{5-7} 
0.40 & 11.666 - 0.0171i & 2 &  &  & 11.687 - 0.114i & 2\tabularnewline
\cline{1-3} \cline{5-7} 
\end{tabular}\protect\caption{Resonances for the $xx\bar{b}\bar{b}$ system(left) and for the $xb\bar{x}\bar{b}$
(right)}
\label{Table:resonances}
\end{table}

\subsection{Crypto-exotic channels}

We also study the crypto-exotic $qQ\bar{q}\bar{Q}$ system, for $m_{Q}=m_{b}=4.7\,\mbox{GeV}$
and $m_{q}=m_{x}$ varying from $0.4$ to $1.3\,\mbox{GeV}$ . We
don't find any bound states and only find resonances for $m_{x}=0.40\,\mbox{GeV}$
and $m_{x}=0.70\,\mbox{GeV}$ . Their energies are displayed in Table
\ref{Table:resonances}.

\section{Conclusion}

An unitarized method to compute the meson-meson scattering was developed.
With it we were able to find bound states and resonances for the $0^{+}$
$xx\bar{b}\bar{b}$ system. For the $xb\bar{x}\bar{b}$ system, only
resonances were found and for sufficiently small $m_{x}$. Refinements
should be easy to include in this model.

Our results however, seems to disagree with lattice results, because
the bound state for exotic system has $S_{12}=0$ and so is a scalar
isotriplet, but, according to \cite{Wagner:2011ev} such a system
should be repulsive. More work is needed to understand the source
of this discrepancy and whether it is problem with the potential model
or with the approach itself. 

Marco Cardoso is supported by FCT under the contract SFRH/BPD/73140/2010. 

\bibliographystyle{unsrtnt}
\bibliography{bib}

\end{document}